\null
%\nopagenumbers
\magnification=\magstep1
\hsize=16.2truecm
\vsize=23.5truecm
\voffset=0\baselineskip
\parindent=1truecm
\tolerance=10000
\baselineskip=17pt
\voffset=0\baselineskip
\parskip=0.5truecm
%\par
 
\def\al{\alpha}
\def\be{\beta}
\def\ga{\gamma}
\def\c{\nabla}

\def\d{\partial}

\def\de{\delta}
\def\e{\epsilon}
\def\f{\phi}
\def\De{\Delta}

\def\gg{\tilde g}

\def\I{{\bf I}}
\def\la{\lambda}
\def\La{\Lambda}

\def\oo{\over}

\def\si{\sigma}

\def\ta{\tau}
\def\~{\tilde}
\def\^{\hat}
\font\twentyninerm   = lcmss8  scaled 3583 % roman
\noindent
\line{\bf KERR-SCHILD METRICS REVISITED II.\hfil}
\noindent
{\bf THE COMPLETE VACUUM SOLUTION}\footnote{$^\dagger$}{Research
supported by OTKA fund no. 1826}\hfil
\vskip .1in
\noindent
{\sl L\'aszl\'o \'A. Gergely
\footnote{$^\ddagger$}{Present address:
{\it Research Group on Laser Physics of the
Hungarian Academy of Sciences, H-6720 Szeged, D\'om t\'er 9,
Hungary}}
and Zolt\'an Perj\'es}\hfil
\vskip .0in
\noindent
\line{Central Research Institute for Physics\hfil}
\line{H-1525 Budapest 114, P.O.Box 49, Hungary\hfil}
\vskip.05in
\noindent
{\bf ABSTRACT}
\midinsert
\baselineskip=12pt plus.1pt
   The complete solution of Einstein's gravitational equations
with a vacuum-vacuum Kerr-Schild pencil of metrics
$g_{ab}+V l_al_b$ is obtained. Our
result generalizes the solution of the Kerr-Schild problem with a
flat metric $g_{ab}$ (represented by the Kerr theorem)
to the case when $g_{ab}$ is the metric of a curved space-time.
 
\endinsert
\baselineskip=17pt
\voffset=0\baselineskip
\par
{\sl PACS classification:} 04.20Jb
\par
 
{\bf 1. INTRODUCTION}
\smallskip
  The Kerr-Schild pencil of metrics in general relativity has the
form$^1$
$$\tilde g_{ab}=g_{ab}+V l_a l_b          \eqno(1.1)$$
where the metrics $g_{ab}$ and $\gg_{ab}$ are both Lorentzian, the
vector $l$ is null with respect to both metrics, and $V$ is a
scalar function.
 
   In part {\bf I} of this series$^2$, we have established some
generic properties of the Kerr-Schild pencil for which both
the parent $g_{ab}$ and $\gg_{ab}$ are vacuum metrics.
The vector $l$ is then
tangent to a null geodesic congruence.
 
   In this paper we present the complete solution of the
vacuum Kerr-Schild problem. The structure of our solution is
as follows. A Kerr-Schild space-time is
characterized by a real {\it deformation parameter $\eta$}.
The deformation parameter vanishes for Kerr-Schild
\vfill\eject\noindent
space-times
the parent of which is a Minkowski space-time. By our theorem in
{\bf I}, the field quantities are severely restricted unless the
$sin\eta$ assumes either of the exceptional values
$0,\pm1, \pm\sqrt2/2$. Here, we obtain all the metrics
with arbitrary values of the deformation parameter. These turn out
to be K\'ota-Perj\'es metrics. For the exceptional values of $\eta$,
we establish that (a) the class with $\eta=0$ is algebraically
special, (b) the values $sin\eta=\pm1$ can occur only in
automorphisms of
Minkowski space-time, and (c) the class with $sin\eta=\pm1/\sqrt2$
contains the remaining K\'ota-Perj\'es metrics.
Our results, as an important implication, dash
the hopes for a complex-analytic description of space-time within
the framework of Kerr-Schild theory.
 
    In {\bf I},
the vacuum Kerr-Schild equations have been written down in a
Newman-Penrose (NP) form$^3$, choosing $l$ a vector of the
null tetrad.
The geodesic condition reads:
  $$ \kappa=0\ . \eqno(1.2)$$
Following {\bf I}, we adopt the gauge with
$$\e=0\ ,\qquad  \pi=\al+\bar\be \  .\eqno(1.3)$$
We have been able to integrate a closed subset of the `radial' field
equations$^3$ containing derivatives in the direction of
the affine parameter $r$:
  $$D=l^a\c_a=\partial/\partial r . \eqno(1.4)$$
This has yielded the spin coefficient quantities
$$\rho=-{1\oo 2r}(1+cos\eta\ C)    \ ,     \qquad
  \si=-{sin\eta\oo 2rC}\                \eqno(1.5)$$
$$\Psi_0=-{sin2\eta\oo 4r^2}        \ . \eqno(1.6)$$
Here we introduce the complex phase factor
$$C={r^{cos\eta}-iB\oo r^{cos\eta}+iB} \eqno(1.7)$$
with the properties
$$\bar C={1\oo C},\qquad
DC={cos\eta\oo2r}(1-C^2)\ .\eqno(1.8)$$
The real integration functions $B$ and $\eta$
may depend on
the coordinates $(x^1,x^2,x^3)$. A further integration function
has been
eliminated from $\rho$ by the appropriate choice of the origin of
the affine parameter. When $r\ge0$, the real potential can be written
in the form
$$V= V_0{r^{cos\eta}\oo r^{2cos\eta}+B^2}\ .\eqno (1.9)$$
The complex tetrad vector $m$ has the $r$ dependence
$$m={1\oo2B}\Bigl(1-{1\oo C}\Bigr)
\Bigl[iQ_1^j r^{cos\eta-\sin\eta-1\oo2}-Q_2^j r^{cos\eta+\sin\eta-1\oo2}
\Bigr]\,{\d\oo\d x^j}\qquad j=1,2,3 \ \eqno(1.10)$$
where $V_0,Q_1^j$ and $Q_2^j$ are real integration functions.
With this choice of the tetrad, the spacelike and null rotations
have been completely used up (cf. {\I}).
 
  The real parameters $\eta$ and $B$ tune the amount of divergence
and rotation of the null congruence with tangent $l$, respectively.
$B$ occurs in the phase factor $C$, giving rise to the
imaginary part of the spin coefficient $\rho$. When
$B=0$, we have $C=1$, and the congruence
is curl-free. Similarly, for large values of the affine parameter
$r$, $C$ approaches the unit value, and the
rotation dies out. The parameter $\eta$ regulates the shear.
For $\eta=0$ or $\eta=180^\circ$, the congruence
is shear-free. When both $B=0$ and
$\eta=0$, the rays are exactly spherical: $\rho=-1/r$.
The rays become cylindrical, $\rho=-1/2r$, for
$\eta=90^\circ$. When $\eta=180^\circ$, there is no expansion.
We have shewn in \I\ that the general shearing class
does not contain the shear-free case as a smooth limit.
In what follows, we shall consider the generic case.
 
   The field equations in the NP form may
be grouped in three sets. The first set of equations is a
coupled system of linear homogeneous equations for the affine
parameter dependence of the quantities $\pi,\ta,\al,\be,\Psi_1$
and for their complex conjugates:
$$\eqalign{\twentyninerm T&HE\cr&\Psi_1\cr
E&QS.\cr}\qquad\qquad\qquad\qquad \eqalign{
D\ta&=\rho(\ta+\bar\pi)+\si(\bar\ta+\pi)+\Psi_1  \cr
D\pi&=2\rho\pi+2\bar\si\bar\pi+\bar\Psi_1        \cr
D\al&=\rho(\pi+\al)+\bar\si(\bar\pi-\bar\al)     \cr
D\be&=\bar\rho\be+\si(2\pi-\bar\be)+\Psi_1       \cr
D\Psi_1&=4\rho\Psi_1+(\pi-4\al)\Psi_0.   \cr}
\qquad\qquad\qquad\qquad\eqalign{
&(1.11a) \cr
&(1.11b) \cr
&(1.11c) \cr
&(1.11d) \cr
&(1.11e) \cr}$$
This set, to be called the $\Psi_1$ equations, can be obtained by
using (\I\ 3.13) in (NP 4.2.c), (NP 4.2.d), (NP 4.2.e), and
the first Bianchi equation (NP 4.5). It is complemented by two
algebraic equations, linear in the unknown functions. The latter
conditions will be employed in Sec. 2 to get a quartet of coupled
equations for the spin coefficients $\ta,\pi$ and their complex
conjugates. The general solution of the $\Psi_1$ set is given in
Sec. 2.
 
   The second system of the field equations,
for the affine parameter dependence of the fields
$\la,\mu$ and $\Psi_2$, is linear and inhomogeneous:
$$\eqalign{\twentyninerm T&HE\cr&\Psi_2\cr
E&QS.\cr}\qquad\qquad\qquad\qquad \eqalign{
&D\la=\rho\la+\bar\si\mu
+\bigl(\bar\de\pi+2\al\pi\bigr)\cr
&D\mu=\bar\rho\mu+\si\la+\Psi_2
+\bigl(\de\pi+2\be\pi\bigr)\cr
&D\Psi_2=3\rho\Psi_2-\la\Psi_0
+\bigl(\bar\de\Psi_1+2\bar\be\Psi_1\bigr)\ .\cr}
\qquad\qquad\qquad\eqalign{
&(1.12a) \cr
&(1.12b) \cr
&(1.12c) \cr}$$
This set will be named the $\Psi_2$
equations. The complex conjugate fields $\bar\la$, $\bar\mu$ and
$\bar\Psi_2$ are decoupled here, and
the source terms in the brackets are provided nonlinearly by the solution of the
$\Psi_1$ system.  In Sec. 3 we find the general solution of the
homogeneous part of the $\Psi_2$ system.
 
  Thus the affine-parameter dependence of the field quantities is
essentially governed by the $\Psi_1$ and the $\Psi_2$ equations.
The third (`nonradial'$^3$) set of equations yields some
constraints and also
the integrals of the remaining field quantities $\ga, \nu, \Psi_3$
and $\Psi_4$. In Section 4, we obtain
the metrics with the trivial solution of the $\Psi_1$ system
and with the general solution of the $\Psi_2$ system.
We get a three-parameter
pencil for which both the parent metric and the image
of the Kerr-Schild map are K\'ota-Perj\'es metrics.
 
  In Sec. 5 we prove that there are no other metrics in the class
with arbitrary values of $\eta$ than those given in Section 4.
The exceptional values of the parameter $\eta$ are considered in
Section 6. We find that all K\'ota-Perj\'es metrics$^7$ are explicit
cases of Kerr-Schild space-times, either with a real deformation
parameter or with $\sin\eta=1/\sqrt2$. This paper is an
expanded version of our preprint on the homogeneous integrals$^4$.
 
%\medskip
{\bf 2. THE $\Psi_1$ EQUATIONS}
\smallskip
  The $\Psi_1$ set of field equations [Eqs.(1.11)]  is linear and
homogeneous for the quantities $\ta,\pi,\al,\be,\Psi_1$ and their
complex conjugates.
Two linear algebraic relations among these quantities follow from
the adopted gauge, $\pi-\al-\bar\be=0$, and from the Kerr-Schild
condition (\I\ 4.1);
$$\Psi_1=\rho\bar\pi+\si(\pi-4\al)+(\rho-\bar\rho)\ta \eqno(2.1)$$
and Eq. (\I\ 4.4) yields the second algebraic constraint:
$$4\si\al=\si(3\pi-\bar\ta)+\Bigl(\rho-{\Psi_0\oo2\si}\Bigr)\bar\pi
+\Bigl({\Psi_0\oo2\si}-\bar\rho\Bigr)\ta\ .\eqno(2.2)$$
Using the Kerr-Schild constraint (2.1) for eliminating $\Psi_1$
in (1.11a) and (1.11b), we
get the four coupled equations for $\pi,\ta$;
$$\eqalign{&
D\ta=(2\rho-{\Psi_0\oo2\si})\ta+\si(2\bar\ta-\pi)
                +(\rho+{\Psi_0\oo2\si})\bar\pi\cr
& D\pi=(2\rho+{\Psi_0\oo2\bar\si})\pi+\bar\si\ta
        + (\bar\rho-{\Psi_0\oo2\bar\si})\bar\ta} \eqno(2.3)  $$
with their complex conjugates.
Eqs. (1.11c), (1.11d) and (1.11e) are a consequence of the algebraic
relations and Eqs. (2.3).
The solution of Eqs. (2.3) determines the solution of the complete
system (1.11), (2.1) and (2.2).
 
  We solve Eqs. (2.3) by separating an overall power of $r$ from
the unknown functions:
$$\pi={1\oo r^{p/2}}\pi^o\ ,\qquad\ta={1\oo r^{p/2}}\ta^o\ .\eqno(2.4)$$
Here we consider $\pi^o$ and $\ta^o$ to be functions of the
complex phase factor $C$
as the new independent variable. Thus the functions $\pi^o$ and
$\ta^o$ satisfy the equations
$$\eqalign{&cos\eta(C^2-1) \dot\ta^o=(2-p+3cos\eta
C)\ta^o+2{sin\eta\oo C}\bar\ta^o-{sin\eta\oo C}\pi^o+\bar\pi^o\cr
&cos\eta(C^2-1)\dot{\bar\pi}^o=(1+2cos\eta C)\ta^o+{sin\eta\oo
C}\bar\ta^o+ (2-p+2{cos\eta \oo C}-cos\eta C)\bar\pi^o}
\eqno(2.5) $$
and the equations for the complex conjugates.
We denote the derivative with respect to $C$ by a dot.
 
   We seek the solution in the form of the finite series in $C$:
$$\pi^o=\sum_{-n}^n P_k C^k,\qquad\ta^o=\sum_{-n}^n T_k C^k\
.\eqno(2.6)$$
Substituting in (2.5) and collecting the like powers of $C$, we
obtain a set
of homogeneous algebraic equations for the coefficients $P_k$ and
$T_k$. Solutions exist when the determinant of the algebraic
equations vanishes. For $n=3$, the determinant is
       $${\cal D}\oo 9cos^2\eta-sin^2\eta\, (p-4)^2\eqno(2.7)$$
where ${\cal D}$, a $12^{th}$
order polynomial in $p$, can be factorized in the form
$$\eqalign{
{\cal D}=&(p^2-6cos\eta \,p-6p+10cos^2\eta+18cos\eta+8)\times\cr
         &(p^2+6cos\eta \,p-6p+10cos^2\eta-18cos\eta+8)\times\cr
         &(p^2-2cos\eta \,p-6p+ 2cos^2\eta+ 6cos\eta+8)\times\cr
         &(p^2+2cos\eta \,p-6p+ 2cos^2\eta- 6cos\eta+8)\times\cr
         &(p^2-2cos\eta \,p-2p+ 2cos^2\eta+ 2cos\eta  )\times\cr
         &(p^2+2cos\eta \,p-2p+ 2cos^2\eta- 2cos\eta  ) \ .}
                                            \eqno(2.8)$$
Thus solutions exist in the twelve cases
$$\eqalign{p&=\pm3cos\eta\pm sin\eta+3\cr
           p&=\pm cos\eta\pm sin\eta+3\cr
           p&=\pm cos\eta\pm sin\eta+1\ .}\eqno(2.9)$$
In each case, the coefficient equations are satisfied by choosing
the overall factor of the solution $P_{-3}$ either real or pure
imaginary. The correct choice of $P_{-3}$ can be found in Table 1.
The explicit forms of the solutions $P_k$ and $T_k$ are given in
Tables 2 and 3. In each case the solution is a rational
expression in $cos\eta$ and $sin\eta$ which, however, is elaborate
for some values of $k$.
The functions $\al$ and $\Psi_1$ may be given
in a similar representation:
$\al={1\oo r^{p/2}}\sum_{-3}^3 A_k C^k,\qquad
\Psi_1={1\oo r^{p/2+1}}\sum_{-2}^4 \psi_k C^k\ .$
The coefficients $A_k$ and $\psi_k$ are determined by the linear
algebraic equations (2.1) and (2.2).
\parskip=0.8truecm
 
  Returning to the affine parameter $r$ as the independent
variable, the solutions for $\tau$ and $\pi$ with
         $ p=\pm 3cos\eta\pm sin\eta+3$ and with
         $ p= \pm cos\eta\pm sin\eta+3$
respectively turn out to be pairwise equal up to a constant
factor. Similarly, solutions with the opposite signs of $cos\eta$
in $p$ will pairwise coincide. Thus, in accordance with the
general theory of linear differential equations$^6$, we are
left with the fundamental solution consisting of the four linearly
independent cases
$$\eqalign{p^{(1)}&= cos\eta+sin\eta+1\cr
           p^{(2)}&= cos\eta-sin\eta+1\cr
           p^{(3)}&= cos\eta+sin\eta+3\cr
           p^{(4)}&= cos\eta-sin\eta+3   \ .}\eqno(2.10)$$
 
  The four fundamental solutions for $\pi$ and $\tau$ are:
$$\eqalign{
\pi^{(1)}=\ {C+1\oo r^{cos\eta+sin\eta+1\oo2}}
&\Bigl(C^{-3}-{cos\eta\oo sin\eta+1}C^{-2}
 -{sin\eta\oo sin\eta+1}C^{-1}                     \cr
&+{5sin^2\eta-4sin\eta+3\oo cos\eta(sin\eta-3)}
+{3sin\eta-1\oo sin\eta+1}{2sin\eta+3\oo sin\eta-3}C
+2{sin\eta\oo cos\eta}C^2\Bigr)                    \cr\cr
\pi^{(2)}=i{C+1\oo r^{cos\eta-sin\eta+1\oo2}}
&\Bigl(C^{-3}+{cos\eta\oo sin\eta-1}C^{-2}
-{sin\eta\oo sin\eta-1}C^{-1}                      \cr
&+{5cos^2\eta-4sin\eta-8\oo cos\eta(sin\eta+3)}
+{6cos^2\eta+7sin\eta-3\oo cos^2\eta-2 sin\eta+2}C
-2{sin\eta\oo cos\eta}C^2\Bigr)                    \cr\cr
\pi^{(3)}=i{C-1\oo r^{cos\eta+sin\eta+3\oo2}}
&(C+1)^2\Bigl(-C^{-3}-{sin\eta+1\oo cos\eta}C^{-2}
+{1\oo sin\eta-1}C^{-1}
-2{sin\eta\oo cos\eta}\Bigr)                       \cr\cr
\pi^{(4)}=\ {C-1\oo r^{cos\eta-sin\eta+3\oo2}}
&(C+1)^2\Bigl(-C^{-3}-{cos\eta\oo sin\eta+1}C^{-2}
-{1\oo sin\eta+1}C^{-1}
+2{sin\eta\oo cos\eta}\Bigr)                       \cr\cr
                                    }\eqno(2.11)$$
$$\eqalign{
\tau^{(1)}=\ {C+1\oo r^{cos\eta+sin\eta+1\oo2}}
&\Bigl({sin\eta\oo cos\eta}C^{-3}-{sin\eta\oo sin\eta+1}C^{-2}
 -{sin^2\eta+8sin\eta+3\oo cos\eta(sin\eta-3)}{sin\eta-1\oo
sin\eta+1}C^{-1}   \cr
&-3{sin\eta\oo sin\eta+1}
+{2sin\eta-3\oo cos\eta}C
+2C^2\Bigr)                    \cr\cr
\tau^{(2)}=i{C+1\oo r^{cos\eta-sin\eta+1\oo2}}
&\Bigl({sin\eta\oo cos\eta}C^{-3}+{sin\eta\oo sin\eta-1}C^{-2}
 -{sin\eta cos^2\eta-7cos^2\eta+4sin\eta+4\oo
cos\eta(2sin\eta-2-cos^2\eta)}C^{-1}   \cr
&+3{sin\eta\oo sin\eta-1}
+{2sin\eta+3\oo cos\eta}C
-2C^2\Bigr)                    \cr\cr
\tau^{(3)}=i{C-1\oo r^{cos\eta+sin\eta+3\oo2}}
&(C+1)^2\Bigl(-{sin\eta\oo cos\eta}C^{-3}+{sin\eta\oo
sin\eta-1}C^{-2} -{2sin\eta+1\oo cos\eta}C^{-1}
-2\Bigr)                       \cr\cr
\tau^{(4)}=\ {C-1\oo r^{cos\eta-sin\eta+3\oo2}}
&(C+1)^2\Bigl(-{sin\eta\oo cos\eta}C^{-3}-{sin\eta\oo
sin\eta+1}C^{-2} -{2sin\eta-1\oo cos\eta}C^{-1}
+2\Bigr)                 \ .      \cr
                                    }\eqno(2.12)$$
\vfill\eject\noindent
Hence we get the fundamental solutions for $\Psi_1$:
$$\eqalign{
\Psi_1^{(1)} = \ {C+1\over2r^{cos\eta+sin\eta+3\over2}} \Bigl(
  sin&\eta C^{-2}
+ (sin\eta - 5) {sin\eta\over cos\eta}C^{-1}
+ {16 sin^2\eta - 15 sin\eta +  9\over sin\eta - 3}\cr
&-{2 sin^2\eta - 3 sin\eta - 3\over cos\eta} C
 - 3 (sin\eta - 1) C^2
 - 3 cos\eta C^3  \Bigr)\cr\cr
\Psi_1^{(2)} =
 i{C+1\over 2r^{cos\eta-sin\eta+3\over2}} \Bigl(
 sin&\eta C^{-2}
- (sin\eta + 5) {sin\eta\over cos\eta}C^{-1}
 + {16 sin^2\eta + 15 sin\eta + 9\over sin\eta+3}\cr
&+{2 sin^2\eta + 3 sin\eta - 3\over cos\eta} C
- 3 (sin\eta + 1) C^2
+ 3 cos\eta C^3 \Bigr)\cr\cr
\Psi_1^{(3)} = i\ {(C-1)(C+1)^2\over2r^{cos\eta+sin\eta+5\over2}}&\Bigl(
   3 cos\eta C
 + 3(sin\eta + 1)
 - (sin\eta - 3) {sin\eta\over cos\eta}C^{-1}
 - sin\eta C^{-2}
\Bigr)\cr\cr
\Psi_1^{(4)} =
\ {(C-1)(C+ 1)^2 \over2r^{cos\eta-sin\eta+5\over2}}&
 \Bigl(
 - 3 cos\eta C
 + 3 (sin\eta - 1)
 + (sin\eta + 3) {sin\eta\over cos\eta} C^{-1}
 - sin\eta C^{-2}
\Bigr) \ .\cr}\eqno(2.13)$$
The cases labeled (3) and (4) are found to satisfy the relation
$\tau=2\be$.
 
  The general solution of the $\Psi_1$ system is given by
constant linear combinations with $real$ coefficients of the four
fundamental
solution vectors. This is not to say, though, that the general
solution will satisfy the complete set of Einstein's vacuum
equations. In fact, one can exclude some of the linear
combinations, without solving the rest of the vacuum
equations, merely by looking at the limit $\eta\to0$.
When $\eta=0$, the parent metric is algebraically special.
Certainly, the trivial solution of the $\Psi_1$ system is
well-behaved in this limit. However, it is easy to see that none
of the four
fundamental solutions (2.13) for $\Psi_1$ vanishes in the limit
considered. One can take linear combinations proportional to
$sin\eta$, but these solutions of the $\Psi_1$ system are trivial
for $\eta=0$. Nontrivial solutions can also be constructed.
Remembering the relation (1.7) between the
variables $C$ and $r$, one finds that the four solution vectors
pairwise coincide in the limit $\eta\to0$. The solutions with
$p=cos\eta\pm sin\eta+1$ coincide, up to a constant multiplier,
with the solutions
with $p=3cos\eta\pm sin\eta+3$. Thus one can form the two linear
combinations
$$\Psi_1^{(+)}= B\Psi_1^{(3)}-\Psi_1^{(1)}\ , \eqno(2.14)$$
$$\Psi_1^{(-)}= B\Psi_1^{(4)}-\Psi_1^{(2)} \eqno(2.15)$$
with the correct limiting behavior.
%Application of L'Hospital's
%rule shows that the fields $\Psi_1^{(\pm)}$ vanish when
%$\eta\to0$.
 
\bigskip
{\bf 3. THE $\Psi_2$ EQUATIONS}
 
  The $\Psi_2$ system (1.12) is a first-order linear and
inhomogeneous set of equations for
the field quantities $\la,\mu$ and $\Psi_2$. The source terms are
supplied by derivatives and quadratic algebraic
expressions of the solutions of the $\Psi_1$ equations.
Let us solve first the linear homogeneous system
$$\eqalignno{&D\la=\rho\la+\bar\si\mu
                               &(3.1a)\cr
&D\mu=\bar\rho\mu+\si\la+\Psi_2
                           &(3.1b)\cr
&D\Psi_2=3\rho\Psi_2-\la\Psi_0
                                         &(3.1c)}$$
which amounts to taking the trivial solution of the $\Psi_1$ Eqs.,
$$\ta=\pi=\Psi_1=\al=\be=0\ .\eqno(3.2)$$
 
   We can decouple
equation (3.1c) for $\Psi_2$ from the rest by considering
first the homogeneous equation
$$D\Psi_2=3\rho\Psi_2.\eqno(3.3)$$
The solution is
$$\Psi_2=\Psi_2^o\Biggl[
{r^{cos\eta-1\oo2}\oo r^{cos\eta}+iB}\Biggr]^3 \eqno(3.4)$$
where $\Psi_2^o$ is a constant of integration. We now turn to
the method of variation of constants, and allow for the
possibility that
$\Psi_2^o$ depends on $r$. Introducing the new functions
$\la^o$ and $\mu^o$
by putting
$$\la=\la^o{r^{3cos\eta+1\oo2}\oo(r^{cos\eta}+iB)^3},\quad \quad
\mu=\mu^o{
r^{3cos\eta+3\oo2}\oo(r^{cos\eta}+iB)^2(r^{cos\eta}-iB)},\eqno(3.5)$$
we can write the system (3.1) as
$$\eqalignno{\textstyle
D\la^o&={1\oo
r}(cos\eta\,C-1)\la^o-{1\oo2}sin\eta\,\mu^o&(3.6.a)\cr
D\mu^o&={1\oo r}C cos\eta\,x
-{1\oo2r^2}(4r\mu^o+sin\eta\,\la^o)&(3.6.b)\cr
D\Psi_2^o&={1\oo2}sin\eta\,cos\eta\,\la^o&(3.6.c)}$$
where we denote
$$x=\mu^o+{1\oo cos\eta}{\Psi_2^o\oo r^2} \ .\eqno(3.7)$$
 
  Comparing Eqs. (3.6.b) and (3.6.c), we find that the function
$x$ satisfies the uncoupled equation
$$Dx=-{2\oo r}x+{cos\eta\oo r}Cx. \eqno(3.8)$$
This has the solution
$$x=x^o{(r^{cos\eta}+iB)^2\oo r^{cos\eta+2}}\eqno(3.9)$$
where $x^o$ is a function of integration. Thus Eqs. (3.6.a) and
(3.6.b) become a pair
of coupled inhomogeneous linear equations for $\la^\circ$ and
$\mu^\circ$ with the driving term ${1\oo r}C cos\eta\,x$.
 
  Let us introduce the functions $L$ and $N$ of the variable
$C$ by
$$\la^\circ={1\oo1-C^2}{L\oo r},\qquad
  \mu^\circ={1\oo1-C^2}{N\oo r^2}\ . \eqno(3.10)$$
$L$ and $N$ satisfy the inhomogeneous equations
$$\eqalignno{\textstyle
N&={C^2-1\oo tan\eta} \dot L &(3.11)\cr
\ddot L-\Bigl({tan\eta \oo C^2-1}\Bigr)^2
L&=-A_0tan^3\eta{C\oo(C^2-1)^2} &(3.12)}$$
where $A_0={8iB\oo tan^2\eta}x^o$. The homogeneous part of
Eq. (3.12) can be transformed to a Riccati equation by the
substitution
$L=L_oe^{\int ydC}$ where $L_o$ is a constant. A particular
solution for $y$ is given by
$$y_1={1\oo2cos\eta}\Bigl({1+cos\eta\oo 1+C}+{1-cos\eta\oo
1-C}\Bigr)\ .\eqno(3.13)$$
Substituting next $y=y_1+z$, the Riccati equation becomes a
Bernoulli equation for $z$ with the general solution
$$z=-{2\oo cos\eta(1-C^2)(1+L_1r)}\ .\eqno(3.14)$$
Here $L_1$ is a constant of integration. Thus the
solution of the homogeneous part of Eq. (3.12) is
$$L_{hom}=L_o(C^2-1)^{cos\eta-1\oo 2cos\eta}
     [(C-1)^{1\oo cos\eta}+(iB)^{1\oo cos\eta}L_1(C+1)^{1\oo
cos\eta}]\eqno(3.15)$$
and the solution of the inhomogeneous equation is$^6$:
$$L=A_0tan\eta\ C+(1-C^2)^{1\oo2}[A_1r^{1\oo2}+
                           A_2r^{-{1\oo2}}]\eqno(3.16)$$
where $L_o$, $A_0,A_1$ and $A_2$ are real functions of
integration.
 
Hence we get the solution of the homogeneous $\Psi_2$ system:
$$ \la=-sin\eta{C-1\oo8cos\eta B}\Bigl\{{2i cos\eta}(C+1)\Bigl(
            A_1+{A_2\oo r}\Bigr)-A_0r^{cos\eta-1\oo2}{C\oo B}
                                   \Bigr\} \eqno(3.17)$$
 
$$\eqalign{
   \mu=-i{C-1\oo8 B^2C}\Bigl\{2B(C+1)[
         &cos\eta C(A_1+{A_2\oo r})-(A_1-{A_2\oo r})]
        -A_0{(C-1)^2\oo B}r^{3cos\eta-1\oo2}
                    \Bigr\}}\eqno(3.18)$$
 
$$\eqalign{
\Psi_2={C-1\oo4cos\eta\ B^2}&\Bigl\{-{1\oo4}A_0r^{cos\eta-3\oo2}
           (2cos^2\eta\ C^2-cos^2\eta-1) \cr
         &+iA_1{cos^2\eta\oo r}B(cos\eta C-1)(C+1)      \cr
         &+iA_2{cos^2\eta\oo r^2}B(cos\eta C+1)(C+1)
                    \Bigr\}\ .}\eqno(3.19)$$

\bigskip
{\bf 4. THE HOMOGENEOUS METRICS}
 
  In this section we carry through the solution procedure of the
field equations for the general solution (3.17)-(3.19) of the
homogeneous $\Psi_2$ system.
 
  From the equation (NP 4.2.l)
$$\Psi_2=\mu\rho-\la\si+\ga(\rho-\bar\rho)\eqno(4.1)$$
we obtain $\ga$ algebraically:
$$\eqalign{\ga={1\oo4B}&\Bigl\{
     {r^{cos\eta-1\oo2}\oo2Bcos\eta}A_0(1-cos\eta C)(1-C)
                                     \cr
      &-iA_1{1\oo cos\eta}[sin^2\eta+(1-cos\eta C)^2]
       -iA_2{cos\eta\oo r}(C^2-1)\Bigr\}
                                                \ .}\eqno(4.2)$$
 
 We now find the $r$ dependence of the tetrad vector $n$.
This can be done by applying the commutator $[\De,\de]$
to the each of coordinates $r$ and $x^j$:
$$n=n^0 \partial/\partial r +n^j \partial/\partial x^j \eqno(4.3)$$
where $n^j$ are functions independent of $r$ and
$$\eqalignno{n^0=2Re\Bigl\{
&-A_0{i\oo 2B cos\eta}{r^{{cos\eta+1\oo2}}\oo r^{cos\eta}+iB}
               \cr
&+A_1{i r\oo 2Bcos\eta}\Bigl(1-cos\eta+{2iBcos\eta \oo
r^{cos\eta}+iB}\Bigr)
 -A_2{1\oo r^{cos\eta}+iB} \Bigr\}-G   \ .&(4.4)}$$
The real integration function $G$ is determined by the action of
the commutator $[\bar\de,\de]$ on the coordinate $r$ :
$$ G=-{cos\eta+1\oo 2Bcos\eta} Im A_2 \eqno(4.5)$$
 
  The integration functions are severely restricted by the NP
equations involving the $\De$ derivative:
$$\eqalignno{(NP4.5)\ \
\qquad\De\Psi_0=(4\ga-\mu)\Psi_0+3\si\Psi_2
\quad&\longrightarrow A_0=Re A_1=Im A_2=\De cos\eta=0 \cr
(NP 4.2.p)\quad -\De\si=\mu\si+\bar\la\rho-(3\ga-\bar\ga)\si
\quad &\longrightarrow Im A_1=\De B=0\ .&(4.6)\cr}$$
Hence $B$ and $\eta$ are constants.
The commutator $[\de,\De]$ applied to the
coordinate $r$, together with (NP 4.2.o),
$$\quad \de\ga=-\si\nu $$
yields
$$\de A_2=\nu=0\ .\eqno(4.7a)$$
Furthermore,
$$\eqalignno{(NP 4.2.i)\qquad\qquad\qquad\qquad\qquad D\nu=\Psi_3
\quad&\longrightarrow \Psi_3=0 \cr
(NP 4.2.n)\quad
\de\nu-\De\mu=\mu^2+\la\bar\la+(\ga+\bar\ga)\mu
\quad&\longrightarrow \De A_2=0\ . &(4.7.b)}$$
Finally from the last Kerr-Schild equation
$$\eqalignno{\de(\bar\ta&-\pi)+\bar\de(\ta-\bar\pi)+
{1\oo2}(\rho+\bar\rho)\De (ln\ V)=2 Re\Psi_2 \cr
&-(\ga+\bar\ga)(\rho+\bar\rho)+{1\oo2}(\mu+\bar\mu)({\Psi_0\oo\si}
+\rho-\bar\rho)-(\mu-\bar\mu)(\rho-\bar\rho)\cr
&+\ta(\bar\ta+2\al-3\pi)+\bar\ta(\ta+2\bar\al-3\bar\pi)
+4\pi\bar\pi-\pi(2\bar\al-\bar\pi)
-\bar\pi(2\al-\pi)\ &(4.8)}             $$
we get
$n^j \partial (ln\ V_0)/\partial x^j=0 $.
As a consequence, we have:
$$A_0=A_1=\Psi_3=\nu=0 \quad \quad \quad \eta,B,V_0,A_2 \ are\
real\ numbers \eqno(4.9)$$
Let us denote $A_2=M$.
 
  $\Psi_4$ can be determined algebraically from
$$(NP 4.2.j)
\quad\De\la-\bar\de\mu=(\rho-\bar\rho)\nu-\Psi_4\quad\longrightarrow
\Psi_4=-M^2
sin 2\eta\Biggl[
{r^{cos\eta-1}\oo(r^{cos\eta}+iB)^2}\Biggr]^2 \eqno(4.10)$$
The rest of Eqs. (NP 4.2) and (NP 4.5) are
identities.
 
  The commutators $[\de,\De]$ and $[\bar\de,\de]$ when
applied to $x^j$ give the relations:
$$[N,Q_1]=[N,Q_2]=0 \quad [Q_1,Q_2]=-B cos\eta N \eqno(4.11)$$
where the three-vector $N$ is defined $N=\{n^j\}$. By Eqs. (4.11),
we can adapt the coordinates $x$ and $y$ to the vectors $N$ and
$Q_1$ as follows,
$$N=\partial /\partial x \quad\quad Q_1=\partial /\partial y
\quad\quad Q_2=B cos\eta\  y\ \partial /\partial x +\partial
/\partial z \ .\eqno(4.12)$$
 
  We now employ the completeness relation
$g^{ab}=2l^{(a}n^{b)}-2m^{(a}\bar m^{b)}$ to assemble the inverse
metric:
$$g^{ab}={1\oo H}\left(\matrix
{4M&   H&   0&   0\cr
  H&   -2B^2 cos^2\eta\ y^2 r^{sin\eta-1}&   0&
                               -2B cos\eta\ y\ r^{sin\eta-1}\cr
  0&   0&   -2r^{-sin\eta-1}&   0\cr
  0&   -2B cos\eta\ y\ r^{sin\eta-1}&   0&   -2r^{sin\eta-1}}
  \right) \ .\eqno(4.13)$$
Here $H=r^{cos\eta}+B^2 r^{-cos\eta}$.
 
  Notice that (4.13) is just the K\'ota-Perj\'es$^7$
metric (44). The metric $\tilde g^{ab}$ differs from
$g^{ab}$ only by the value of the parameter $4\tilde
M=4M-V_0$. The curvature components are:
$$\eqalignno{\Psi_0&=-{sin 2\eta\oo 4r^2}\cr
 \Psi_1&=\Psi_3=0  \cr
\Psi_2&=M cos\eta r^{cos\eta-2}
{cos\eta (r^{cos\eta}-iB)+(r^{cos\eta}+iB)\oo (r^{cos\eta}+iB)^3}
\cr
\Psi_4&=-M^2 sin 2\eta \Biggl[{r^{cos\eta-1}\oo
(r^{cos\eta}+iB)^2}\Biggr]^2 \ ,\ &(4.14)}$$
These expressions are well-behaved in the limit $r\to \infty$, and
singular only in the limit $r\to 0$. The space-time is Type I in
the Petrov classification.
 
  To summarize, the Kerr-Schild map in \I\ generates
a K\'ota-Perj\'es metric from a type N vacuum metric.
The image of the map becomes here the parent space-time for the
second Kerr-Schild map to a K\'ota-Perj\'es
space-time with another value of the parameter $M$.

\bigskip
{\bf 5. THE VANISHING OF THE INHOMOGENEOUS TERMS}\par
 
    In this section we show that the presence of source terms
in the $\Psi_2$
equations is incompatible with the nonradial field equations.
To this effect, we prove that $\al=\be=\ta=\pi=\Psi_1=0$.
Our procedure consists in proving the following lemmas:
\item{(i)}
  In the generic case, $\rho-\bar\rho\neq0$, for any real
function $\f$ with $D\f=\de\f=0$, it follows also $\De\f=0$.
\item{(ii)} The integration functions $B,\eta$ and $V_0$ are
constants.
\item{(iii)} The vector $n$ has the radial component
$n^0=-(\mu-\bar\mu)/(\rho-\bar\rho)$.
\item{(iv)} The $\De$ operator, when it acts on any of the
quantities $\rho,\si$ or $\Psi_0$, can be written as $\De=n^0D$.
\item{(v)} There exists a linear equation among the source terms of
the nonradial equations which cannot be satisfied unless all source
terms vanish.
 
We can establish (i) easily by use of the last commutator (NP 4.4):
$$(\bar\de\de-\de\bar\de)\f=[(\bar\mu-\mu)D+(\bar\rho-\rho)\De
 -(\bar\al-\be)\bar\de-(\bar\be-\al)\de]\f\ .\eqno(5.1)$$
When $B=0$, the vector $l$ is curl-free. These fields
will be separately discussed at the end of the section.
The main Theorem in \I\ has already yielded that the $\de$
derivatives of the functions $B, \eta$ and $V_0$ vanish. Hence and
from (i) it follows that $B, \eta$ and $V_0$ are constant.
Thus we have proven (ii) and that
$\rho, \si$ and $\Psi_0$ depend only on the affine parameter $r$.
 
   Choosing $\f=r$ in the commutator (5.1), together with the form
(4.3) of the vector $n$, (iii) follows at once. Next, (iv) is a
straightforward consequence of the fact seen [under (ii)] that the
functions $\rho,\si$ and $\Psi_0$ depend only on $r$.
With $n^0$ given in (iii), we may substitute
$$\De\rightarrow-{\mu-\bar\mu\oo\rho-\bar\rho}D$$ when acting
either on $\rho,\si$ or $\Psi_0$ or any of their complex conjugate
functions.
 
Turning to (v), we observe that the real Kerr-Schild Eq. (4.8),
the complex
Ricci Eqs. (NP 4.2 l,p,q) and the fifth Bianchi identity (NP 4.5),
form a set of linear inhomogeneous algebraic equations for the
spin coefficient quantities $\la, \mu, \ga$ and $\Psi_2$,
$$\eqalignno{
\rho\mu-\si\la-\Psi_2+(\rho-\bar\rho)\ga&=a_1 &(5.2a)\cr
\si\mu-{D\si\oo\bar\rho-\rho}(\bar\mu-\mu)+\rho\bar\la
     -\si(3\ga-\bar\ga)&=a_2                  &(5.2b)\cr
\rho\bar\mu-{D\rho\oo\bar\rho-\rho}(\bar\mu-\mu)+\si\la+\Psi_2
     -\rho(\ga+\bar\ga)&=a_3                  &(5.2c)\cr
\Psi_0\mu-{D\Psi_0\oo\bar\rho-\rho}(\bar\mu-\mu)-3\si\Psi_2
     -4\Psi_0\ga&=a_4                         &(5.2d)\cr
\Bigl[{1\oo2}{\bar\rho+\rho\oo\bar\rho-\rho}\Bigl({1\oo
      r}+\bar\rho+\rho\Bigr)\Bigr](\bar\mu-\mu)
      +{1\oo2}\Bigl({\Psi_0\oo\si}
        +\rho-\bar\rho\Bigr)(\bar\mu+\mu)&\cr
     +\Psi_2+\bar\Psi_2-(\rho+\bar\rho)(\ga+\bar\ga)&=a_5
                                              &(5.2e)
}$$
 The source terms on the right-hand sides are
$$\eqalign{
a_1&=\de\al-\bar\de\be-\al\bar\al-\be\bar\be+2\al\be\cr
a_2&=\de\ta-\ta(\ta+\be-\bar\al)\cr
a_3&=\bar\de\ta+\ta(\bar\be-\al-\bar\ta)\cr
a_4&=\de\Psi_1-(4\ta+2\be)\Psi_1\cr
a_5&=\de(\bar\ta-\pi)+\bar\de(\ta-\bar\pi)-6\pi\bar\pi+2\pi\bar\al
+2\bar\pi\al-2\ta\bar\ta-2\ta\al-2\bar\ta\bar\al
+3\ta\pi+3\bar\ta\bar\pi\ .}      \eqno(5.3)$$
 
   These and the complex conjugate equations are nine real conditions
on eight real unknown functions. Therefore, there exists a linear
combination of the nine equations such that the left hand sides
cancel. This is just the sought-for condition on the source
terms:
$$a_1+{\si\oo\bar\rho}\bar a_2-{a_4\oo3\si}
+\Bigl({\rho\bar\rho+\si\bar\si\oo\bar\rho}-{\Psi_0\oo3\si}\Bigr)
b_1 +\Bigl(\bar\rho-\rho-{4\Psi_0\oo3\si}-{\si\bar\si\oo\bar\rho}
\Bigr)b_2+{3\si\bar\si\oo\bar\rho}\bar b_2=0   \eqno(5.4)$$
where
$$\eqalign{
b_1&={a_1+a_3-\bar a_1-\bar a_3\oo2(\bar\rho-\rho)} \cr
b_2&={1\oo\bar\rho^2-\rho^2}\Big\{\rho{\rho\bar\rho\oo\si\bar\si}
   (-a_5+a_3+\bar a_3)+(3\rho-\bar\rho)(a_1+a_3)
   -{\rho^2\oo\bar\si}\bar a_2-{\rho\bar\rho\oo\si}a_2\cr
 &\qquad\qquad
   +\rho\Bigl[{\rho\bar\rho\oo\si\bar\si}
    \Bigl({1\oo r}+2\rho+2\bar\rho\Bigr)+5\rho-3\bar\rho
    \Bigr]b_1\Bigr\}    \ .}        \eqno(5.5)$$
All the spin coefficient quantities may be expressed, by use
of Eqs. (2.1) and (2.2), in terms of the general solution
$$\pi=\sum_{k=1}^4c_k\pi^{(k)}\qquad\ta=\sum_{k=1}^4c_k\ta^{(k)} \eqno(5.6)$$
and the corresponding complex conjugate quantities. We can take
out the fractional multipliers of the fundamental solutions:
$$ \eqalign{
\pi^{(1)}&=r^{-\sin\eta-1\oo2}\pi_1 \qquad
\pi^{(2)} =r^{ \sin\eta-1\oo2}\pi_2 \qquad
\pi^{(3)} =r^{-\sin\eta-3\oo2}\pi_3 \qquad
\pi^{(4)} =r^{ \sin\eta-3\oo2}\pi_4 \cr
\ta^{(1)}&=r^{-\sin\eta-1\oo2}\ta_1 \qquad
\ta^{(2)} =r^{ \sin\eta-1\oo2}\ta_2 \qquad
\ta^{(3)} =r^{-\sin\eta-3\oo2}\ta_3 \qquad
\ta^{(4)} =r^{ \sin\eta-3\oo2}\ta_4\ . }\eqno(5.7)$$
Thus the affine-parameter dependence of the entries in Eq. (5.4)
is explicitly known, but the detailed form is extremely lengthy.
Since the value of $\eta$ is arbitrary, one can simplify these
computations by choosing some Pythagorean values, {\it e.g.,}
$\sin\eta=3/5$.
 
   We may break up Eq. (5.4) by noting that each coefficient of
the linearly independent functions of $r$ must vanish.
There are two kinds of terms in (5.4); the derivative terms are
linear in $\pi,\ta$ and the complex conjugates and the algebraic
terms are quadratic.
By inspection of (5.7) and the form (1.10) of the $\de$
operator, we find that the
independent powers of $r$ occurring in the derivative terms are:
$\pm sin\eta-2,\pm sin\eta-1,-2$ and $-1$ .
The algebraic terms may have the following powers:
$\pm sin\eta-3,\pm sin\eta-2,\pm sin\eta-1,-3,-2$ and $-1$ .
The powers $r^{sin\eta-3}$ occur only multiplied with the factor
$c_4^2$. Thus, when collecting these terms in (5.4), we can
put $c_1=c_2=c_3=0$. This yields $c_4=0$.
Next, the terms with $r^{-sin\eta-3}$, arising with a factor
$c_3^2$, can be computed in a similar fashion, with the result
that $c_3=0$.
The surviving terms in (5.4) contain the factors
$\pm sin\eta-1$ and $-1$, with derivative and the algebraic terms
mixed. The terms with $sin\eta-1$ do not contain any $c_1$ factor,
hence we can put for these $c_1=0$. As expected, we get $c_2=0$.
Finally, from the terms $-sin\eta-1$ we find $c_1=0$.
 
   We have thus shown that only the trivial solution (Sec. 4) of
the $\Psi_1$ equations satisfies Eq. (5.4).
 
   The {\it curl-free fields} with $B=0$ need to be considered
separately. The computations follow the pattern of the generic
case, and once again only the trivial solution of the $\Psi_1$
system survives. In the class with arbitrary values of $\eta$,
we obtain the Kasner metric$^9$ for $V_0<0$, and
a sign-flipped version of the Kasner metric as
described by McIntosh$^{10}$, for $V_0>0$. Both are
special cases of the K\'ota-Perj\'es metric (\I\ 5.11) for $B=0$.

\medskip
{\bf 6. THE SPACE-TIMES WITH SPECIAL VALUES OF $\eta$}\par
   By our main theorem of \I\  on the vacuum Kerr-Schild
space-times, there exist such exceptional values of the
deformation
parameter $\eta$ for which no restriction follows from Eq. (I.4.7)
for the $\de$ derivatives. To complete our investigation of vacuum
Kerr-Schild space-times, we now consider in turn the metrics with
either of the values $sin\eta=0,\pm1,\pm1/\sqrt{2}$.
 
   (a) When $sin\eta=0$, both $\si$ and $\Psi_0$ vanish,
and $l$ is a principal null vector of the curvature. By the
Goldberg-Sachs theorem$^3$, these parent space-times are
algebraically special, and $\Psi_1=0$.
It then follows from Thompson's Theorem 3.2
that the ensuing space-time is also algebraically special, with
the Kerr-Schild congruence a principal null congruence$^5$.
All the vacuum Kerr-Schild spacetimes generated
from the flat space-time are in this class.
 
   (b) The case with $cos\eta=0$ contains automorphisms of the
Minkowski space-time. With a spatial rotation of the vector $m$, the
$r$-independent phase factor in $\si$ can be removed:
$\rho=\si=-1/2r$. Using these and $\Psi_0=0$ in the $\Psi_1$
equations of \I\ we get $\Psi_1=0$. Then we get from the Bianchi
identities $\Psi_2=\Psi_3=\Psi_4=0$. Calculation shows that
the ensuing space-time is also flat, as expected from the
Goldberg-Sachs theorem for $\si\neq0$.
 
  (c) Case $sin\eta=1/\sqrt{2}=k$ contains the following
K\'ota-Perj\'es metrics:
$$ds^2=-{f^0\oo f}(r^{1-k}dx^2+r^{1+k}dy^2)+2dr
(l_a dx^a)+f(l_a dx^a)^2 \eqno(6.1)$$
with $l=\partial/\partial r$ tangent to the Kerr-Schild
congruence. Metric (53) of Ref. 8 is given by
$$f=\La Re\Bigl\{{x+ir^ky\oo r^k+iB}\Bigr\}, \qquad f^0=\La(x+By)
\eqno(6.2)$$
with $\La$ the pencil parameter and $B$ a real constant.
For metric (66), $B=x/y$ and
$$f=\La {x+by\oo x^2r^{-k}+y^2 r^k}, \qquad f^0=\La(x+by)/y^2\ .\eqno(6.3)$$
Though metric (6.3) is of the Kerr-Schild type, it is {\sl not} a
solution of the vacuum Einstein equations$^8$, and the contrary
claim in Ref. 7 is invalid.
 
    Notice that above we have enlisted all the metrics of Ref. 7.
\item{ }{\it Corollary:} The K\'ota-Perj\'es metrics all belong to the
Kerr-Schild class.
 
\bigskip
{\bf 7. CONCLUDING REMARKS}\par
 
    Our solution of the vacuum Kerr-Schild problem has been made
possible by integration of the $\Psi_1$ system. However, these
lengthy integrals do not satisfy all the field equations.
 
    The vacuum Kerr-Schild pencils generated from a {\sl flat}
space-time are associated with complex surfaces in
three-dimensional homogeneous spaces. This relationship follows
from the Kerr theorem$^{11}$.
It has been known for some time that shear-free null geodesic
congruences do not coexist with the Weyl curvature. Our main
theorem in \I\ has already indicated what we find in this paper
that the variety of {\sl shearing Kerr-Schild} congruences is too
small to allow for the complex-analytic structures of the Kerr
theorem.
 
 {\ \ }
{\bf REFERENCES}
{\frenchspacing
\smallskip
\item{[1]} Kerr, R. P.  and Schild, A., {\it Atti Del
Convegno Sulla Relativit  Generale: Problemi Dell' Energia e Onde
Gravitazionali (Anniversary Volume, Fourth Centenary of Galileo's
Birth)}, G. Barb'ra, Ed. (Firenze, 1965), p. 173
\item{[2]} Gergely, L. \'A. and Perj\'es, Z.: Kerr-Schild metrics
revisited I. The ground state (Paper \I)
\item{[3]} Newman, E. and Penrose, R., J. Math. Phys. {\bf3}, 566
(1962)
\item{[4]} Gergely, L. \'A. and Perj\'es, Z.: Kerr-Schild metrics
revisited II. The homogeneous integrals, KFKI-1993-09/B preprint
\item{[5]} Thompson, A.H., Tensor {\bf17}, 92 (1966)
\item{[6]} Kamke, E.: Differentialgleichungen I., Sec. 8.2.
Teubner, Stuttgart, 1977
\item{[7]} K\'ota, J. and Perj\'es, Z., J. Math. Phys. {\bf13}, 1695
(1972)
\item{[8]} Evans, B., Perj\'es, Z.  and Scott, S., in preparation
\item{[9]} Kasner, E., Amer. J. Math. {\bf 43}, 217 (1921)
\item{[10]} McIntosh, C. B. G., in {\it Relativity Today},
Z. Perj\'es, Ed., Nova Science Publishers, New York (1992), p. 147
\item{[11]} Debney, G. C., Kerr, R. P.  and Schild,
A., J. Math. Phys. {\bf10}, 1842 (1969)
 
\vfill\eject}
 
\par
\smallskip
\def\s{sin\eta}
\def\c{cos\eta}
\newbox\ujstrutbox
\setbox\ujstrutbox=\hbox{\vrule height7pt depth3pt width0pt}
\def\ujstrut{\relax\ifmmode\copy\strutbox\else\unhcopy\strutbox\fi}
\vbox{\tabskip=0pt \offinterlineskip
 \def\tablerule{\noalign{\hrule}}
 \halign to 12.5cm{\strut#& \vrule#\tabskip=5pt plus10pt&
  \hfil#&\hfil \vrule#& \hfil#& \vrule\vrule\vrule#&
  \hfil#& \vrule#&
  \hfil#& \vrule\vrule\vrule#&
  \hfil#& \vrule#&
  \hfil#& \vrule#
   \tabskip=0pt\cr
\tablerule
  &&\omit\hidewidth \hfil \hidewidth&&
    \omit\hidewidth \hfil \hidewidth&&
    \omit\hidewidth \hfil \hidewidth&&
    \omit\hidewidth \hfil \hidewidth&&
    \omit\hidewidth \hfil \hidewidth&&
    \omit\hidewidth \hfil \hidewidth&\cr
  &&\omit\hidewidth   p   \hidewidth&&
    \omit\hidewidth $\ P_{-3}$ \hidewidth&&
    \omit\hidewidth   p   \hidewidth&&
    \omit\hidewidth $\ P_{-3}$ \hidewidth&&
    \omit\hidewidth   p   \hidewidth&&
    \omit\hidewidth $\ P_{-3}$  \hidewidth&\cr
  &&\omit\hidewidth \hfil \hidewidth&&
    \omit\hidewidth \hfil \hidewidth&&
    \omit\hidewidth \hfil \hidewidth&&
    \omit\hidewidth \hfil \hidewidth&&
    \omit\hidewidth \hfil \hidewidth&&
    \omit\hidewidth \hfil \hidewidth&\cr\tablerule
  &&\omit\hidewidth \hfil \hidewidth&&
    \omit\hidewidth \hfil \hidewidth&&
    \omit\hidewidth \hfil \hidewidth&&
    \omit\hidewidth \hfil \hidewidth&&
    \omit\hidewidth \hfil \hidewidth&&
    \omit\hidewidth \hfil \hidewidth&\cr
  &&$ \c+\s+1$&&1&&$ \c+\s+3$&&i&&$ 3\c+\s+3$&&1
                               &\cr
  &&$ \c-\s+1$&&i&&$ \c-\s+3$&&1&&$ 3\c-\s+3$&&i
                               &\cr
  &&$-\c+\s+1$&&i&&$-\c+\s+3$&&1&&$-3\c+\s+3$&&i
                               &\cr
  &&$-\c-\s+1$&&1&&$-\c-\s+3$&&i&&$-3\c-\s+3$&&1
                               &\cr
  &&\omit\hidewidth \hfil \hidewidth&&
    \omit\hidewidth \hfil \hidewidth&&
    \omit\hidewidth \hfil \hidewidth&&
    \omit\hidewidth \hfil \hidewidth&&
    \omit\hidewidth \hfil \hidewidth&&
    \omit\hidewidth \hfil \hidewidth
                               &\cr\tablerule
    \noalign{\smallskip}
    \hfil\cr}}
\par
\smallskip
\ \ \ \ \ Table 1. {\it The twelve solutions for $\tau^o$ and
$\pi^o$} \par
\smallskip
\par
\vfill\eject
 
$$\eqalign{
P_{-3} =&P ,\qquad
P_{-2} = { p - 2 \over cos\eta }P ,\qquad
P_{-1} = { p^2 - 4 p - 2 cos^2\eta + 4 \over  2 cos^2\eta}P,\qquad
P_{0} ={sin\eta\over 2 cos^2\eta}R\bar P \cr
 +&{p^3+(2p^3-12p^2+32p-94) cos^2\eta - 6 p^
2 - 8 p cos^4\eta  + 44 cos^4\eta + 32
\over 2 cos\eta(p^2
cos^2\eta - p^2 - 8 p cos^2\eta + 8 p + 25 cos^2\eta - 16) }P\ ,\cr
 P_{1} =& {3\over 4 cos^2\eta}(p-2)R P
    + {sin\eta \over 4 cos^3\eta (p^2 cos^2\eta - p^2 - 8 p
       cos^2\eta + 8 p + 25 cos\eta ^2 - 16) } \cr
&\quad\times(p^6 - 16 p^5 - 6 p
^4 cos^2\eta + 100 p^4 + 42 p^3 cos^2\eta
 - 304 p^3 + 12 p^2 cos^4\eta\cr&
-72 p^2 cos^2\eta + 448 p^2 - 84 p cos^4\eta - 48 p cos^2\eta - 256 p - 36 cos\eta^
4 + 192 cos^2\eta)
\bar P \ ,\cr
P_{2} =&{3\over 4 cos\eta}R P + {sin\eta
\over 4 cos^2\eta (p^2 cos^2\eta - p^2 - 8 p cos^2\eta + 8 p + 25 cos\eta^
2 - 16) }\cr
&\quad\times(p^5 - 14 p^4 - 2 p^3 cos^2\eta + 68 p^3
- 10 p^2 cos^2\eta - 120 p^2 + 24 p cos^4\eta\cr
&+ 140 p cos^2\eta - 132 cos^4\eta - 200 cos^2\eta + 128)
\bar P \ ,\cr
P_{3} =&2{sin\eta\over cos\eta }\bar P \cr\cr
R =& {p^4 - 12p^3 + 52p^2 - 96 p + 64
+(2p^3-26p^2+84)cos^2\eta+(4p+8)cos^4\eta\over
(8-p)sin^2\eta p+25cos^2\eta-16 }}$$
 
{\it Table 2. The universal coefficients of $\pi$. The fundamental
solution is given by the four
independent values $p=cos\eta\pm sin\eta+1$ and $p=cos\eta\pm sin\eta+3$}
\medskip
\hrule
$$\eqalign{
T_{-3} =&{sin\eta\over cos\eta }P , \qquad
T_{-2} = {sin\eta (p - 2)\over cos^2\eta }P, \qquad
T_{-1} ={sin\eta(p-4)P+3\bar P\over cos^2\eta}R \ ,\cr
T_{0} =&{ sin\eta \over4cos^2\eta}R P
 +{1\over 4 cos^3\eta (p^2 cos^2\eta - p^
2 - 8 p cos^2\eta + 8 p + 25 cos^2\eta - 16) }\cr
&\times( p^5 cos^2\eta - p^5 - 16 p^4 cos^2\eta + 16 p
^4 - 6 p^3 cos^4\eta + 118 p^3 cos^2\eta \cr
&- 100 p^3 + 74 p^2 cos^4\eta - 468
 p^2 cos^2\eta
+ 304 p^2 - 8 p cos^6\eta - 344 p cos^4\eta\cr
&+ 932 p cos^2\eta - 448
 p + 44 cos^6\eta + 456 cos^4\eta - 720 cos^2\eta + 256)
\bar P \ ,\cr\cr
T_{1} =&{p^2 - 5 p - 3 cos^2\eta + 6\over cos^2\eta }\bar P, \qquad
T_{2} = {2 p - 5\over cos\eta }\bar P, \qquad
T_{3} =  2 \bar P}$$

{\it Table 3. The universal coefficients of $\ta$}
\end